\documentclass[12pt]{article}
\usepackage{epsfig}
\input epsf.sty
\newcommand{\be}{\begin{equation}}
\newcommand{\ee}{\end{equation}}
\newcommand{\ba}{\begin{eqnarray}}
\newcommand{\ea}{\end{eqnarray}}
\newcommand{\bi}{\begin{itemize}}
\newcommand{\ei}{\end{itemize}}

\newcommand{\RR}{{\rm I\kern -.2em  R}}

\def\lsi{\raise0.3ex\hbox{$<$\kern-0.75em\raise-1.1ex\hbox{$\sim$}}}
\def\gsi{\raise0.3ex\hbox{$>$\kern-0.75em\raise-1.1ex\hbox{$\sim$}}}

\makeatletter \@addtoreset{equation}{section} \makeatother

\makeatletter
\renewcommand\section{\@startsection {section}{1}{\z@}%
                                   {-5.5ex \@plus -1ex \@minus -.2ex}
                                   {2.3ex \@plus.2ex}%
                                   {\normalfont\large\bfseries}}
\renewcommand\subsection{\@startsection{subsection}{2}{\z@}%
                                     {-3.25ex\@plus -1ex \@minus -.2ex}%
                                     {1.5ex \@plus .2ex}%
                                     {\normalfont\normalsize\bfseries}}
\renewcommand\thesection {\@arabic\c@section}
\renewcommand\thesubsection   {\thesection.\@arabic\c@subsection}
\renewcommand{\@seccntformat}[1]{%
\csname the#1\endcsname.\hspace{1.0em}}
\makeatother

\begin{document}
\begin{center}
{\Large{\bf Cosmological constraints from extra-dimension induced domainwalls\footnote{Prepared for the IVth Marseille International Cosmology Conference, 23-26 June 2003}}}
\medskip

\centerline{\bf Chris P. Korthals Altes }

\centerline{\sl Centre Physique Th\'eorique au CNRS,}
\centerline{\sl Case 907, Campus de Luminy, F13288, Marseille, France}
\end{center}

\abstract{ TeV size extra dimensions introduce domain walls. Such walls are inconsistent
 with CMB anisotropies. Either inflationary dynamics washes them out, or the reheating temperature is lower then the temperature at which the walls start forming networks. As the restoration of the symmetry is  non-perturbative we  have performed 5d lattice simulations which show the occurrence of a cut-off in the fifth dimension as well as symmetry restoration.}

\section{Introduction}\label{sec:intro}

Ever since the Kaluza-Klein papers  the idea of 
extra dimensions has reappeared in  particle physics. One idea is to have 
compact extra dimensions to bring down the Planck scale to say the TeV range~\cite{antoniadis},
thereby drastically changing the unification picture~\cite{dienes}.
In the context of field theory the 5 or higher dimensional bulk contains
at least the gauge fields. It is notoriously difficult to construct models
in which the gauge fields are confined to our 4d world. For quarks such
mechanisms are known since long.

 In this talk I review a simple consequence based on two papers~\cite{laine}: the appearance of a network of    domain walls
in our universe. The anisotropies of the CMB clearly rule out such networks.

In section (\ref{sec:zn}) the occurrence of the global symmetry is explained
and why it appears in spontaneously broken form at low temperature. In section
(\ref{sec:resto}) the restoration of the symmetry at high temperature leads to the phase diagram of the theory with plausibility arguments. Lattice simulations
do confirm the phase diagram. Quite interesting is the cut-off in the extra dimension. If we try to make the lattice mesh smaller, the low temperature
 phase- the one we are supposed to live in- collapses altogether.

But the phase diagram shows that domain walls will appear and in the last section I look at the consequences for cosmology.
   
\section{A global Z(N) symmetry}\label{sec:zn}

Let us limit ourselves to the essentials: we assume an SU(N) gauge theory
in 5d. If there are other fields we will assume them to be confined to 
the 4d world.

The extra coordinate is $y$, and the 4d coordinates are $x$. The length of the 
periodic fifth dimension is $R$: $0<y<R$. The scale $M=R^{-1}$ is supposed to be much smaller than the scale $\Lambda$ of the 4d $SU(N)$ theory.

Long ago 't Hooft~\cite{thooft} noted that gauge transformations need not be periodic in $y$.
Though locally they are genuine gauge transformations, globally they pick up a phase in the center
of the group $SU(N)$. This phase can only be an N'th root of unity, which constitutes a discrete group $Z(N)$.  

In the Hilbert space of physical states such a gauge transformation has a unique effect, {\it only} depending on the discontinuity $\exp{ik{2\pi\over N}}$. This is easy to understand, by comparing two such transformations 
$\Omega_{k}$ and $\Omega_{k}^{\prime}$. They only differ by a regular gauge transform, so have the same effect on a physical state.

So the transformations form a group isomorphic to $Z(N)$ in the physical Hilbert space.

Gauge particles are in the adjoint representation of the gauge group, so do not
feel the centergroup and stay periodic.  Quark fields, being in the fundamental representation, do not stay periodic and break the symmetry. The physical Hilbert space with quarks is not invariant.

From now on we suppose  only particles in Z(N) neutral representations, so that the symmetry is exact. It is a symmetry stemming from the gauge group and the geometry. As the endeavour is strong gravity at the TeV scale, we need a global symmetry that is  robust against strong gravity effects~\cite{colemansuskind}.
 And such is the symmetry at hand.

An important ingredient is an order parameter, that transforms non-trivially
under our group. Obviously a local gauge invariant quantity will be invariant.
One needs a non-local quantity like a string  $P(A_y)$ winding around the $y$-direction. Such a string  is called the spatial Wilson line or Polyakov loop. The Polyakov loop is mathematically defined as an ordered product of gauge phases $\exp{iA_y}$ along the y axis. It transforms under $\Omega_k$ as 
\begin{equation}
P(A_y)\rightarrow \exp{ik{2\pi\over N}}P(A_y).
\end{equation}

The action of $\Omega_k$ on the loop is global. No matter where we put the discontinuity, the loop transforms with a multiplicative factor, which is the centergroup Z(N)~\footnote{ A discontinuity not in the center would have had a local effect on the loop.}. This also summarizes the effect of our Z(N) transformations on physical states: only physical states containing loops will feel
the action of $\Omega_k$.

\section{How is Z(N) realized?}\label{sec:realized}   

In our 5d gauge theory we have a dimensionful coupling constant $g_5$.
The combination $g_5^2M$ is a natural dimensionless combination. 
It appears when we consider the mass scale M to be much smaller than the
$\Lambda$ scale of $SU(N)$. Then we can consider the Kaluza-Klein modes as very
heavy and integrate them out:
\begin{equation}
A_{\mu}(x,y)=\sum_n A_{\mu}(x;n)\exp{in2\pi yM}.
 \end{equation} 

 Let's consider the 5d gauge action and the gradient of our potential in the $y$ direction. If $A_y$ is constant in $y$ such a term is:
\begin{equation}
{1\over{g_5^2}}S(A)= {1\over{g_5^2}M}\int d^4x(\partial_z A_y)^2+... 
\end{equation}

The factor $M$ appears because we integrate out the $y$ coordinate.
The dimensionless combination appears as a 4d coupling $g_4^2=g_5^2M$.

If $g_4$ is a small number we can integrate out the KK modes and the dots are then
the effective potential in  $A_y$. The latter acquires a VEV, and so does the Polyakov loop $P(A_y)$.

So the spontaneously broken phase is realized.

\section{Domainwall tension}\label{sec:tension}

Despite the somewhat arcane nature of the symmetry, its effects in our 4d world
are quite concrete!
Between domains where the Polyakov loop in the compact dimension
takes different centergroup values a domain wall builds up. The profile of such a wall is given by the Polyakov loop, and we have to compute the effective potential already mentioned in the previous section. To do so we set the potential
 $A_y$ equal to its VEV $C$ and small fluctuations $g_4Q_y$:

$$A_y=C+g_4Q_y$$.

The other potentials are considered to be small fluctuations, without background.

By substituting this shift in $A_y$ in the action you see that a classical potential  is absent.

The potential $V(C)$ is entirely a quantum effect, that one gets by integrating out the fluctuations. To one loop order it is just the determinant of the fluctuation matrix in the background of the VEV $C$. This potential reflects the Z(N) symmetry and has minima
in the values of the Polyakov loop $P(C)=\exp{ik{2\pi\over N}}$. 

The walls form in between regions of these centergroup values, and their tension is computed by minimizing the effective action, keeping the values of the loop at beginning
and end fixed in the respective domains. If the domains differ by one unit
in Z(N) one finds for the tension:
\begin{equation}
\sigma_1\sim\int^{2\pi T/N}_0 dC\sqrt{V(C)}={(N-1)\over\sqrt{{g_4^2N}}}M^3.
 \end{equation}
This result~\cite{belyaev} is to be compared with the tension for k units:

\begin{equation}
\sigma_k={k(N-k)\over{N-1}}\sigma_1
 \end{equation}

\noindent so the walls attract~\cite{giovanna}. For $k=O(1)$ they have necessarily  $1/N$ corrections in the large $N$ limit. Thus they constitute a counterexample to the rule that $Z(N)$ neutral systems produce only $1/N^2$ corrections. This rule is based on Feynman diagram analysis. Our calculation involves Feynman diagrams in the background of the soliton forming the wall, invalidating the double line analysis.

The width of the wall is given by ${1\over{\sqrt{g_4^2N}M}}$. The coupling
appears because of the absence of a classical potential. In the minimization procedure we have to balance the kinetic term of order ${1\over {g_4^2}}$ with the quantum potential of order $O(1)$.
 
\begin{figure}[t]

\centerline{\epsfxsize=7cm\hspace*{0cm}\epsfbox{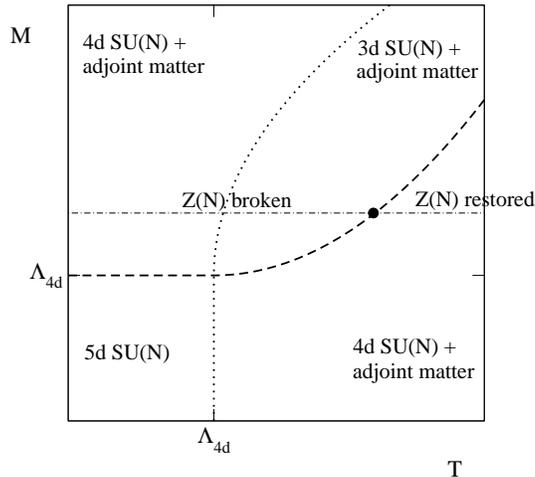}}

\caption[a]{A schematic phase diagram, together with the  
low-energy effective theories in different regions. The Z($N$) symmetry
we have discussed is broken above the dashed line. To the 
right of the dotted line, another Z($N$) symmetry, related to a Polyakov
loop in the Euclidean (finite-$T$) time direction is broken. Our argument follows the horizontal 
dash-dotted line, with the blob indicating the phase transition.}

\end{figure}
\section{Restoration of the Z(N) symmetry}\label{sec:resto}

In fig.1 is shown, what happens when one increases the temperature $T$.
We start on the broken line at $T=0$, in the 4d theory with a small fifth dimension.  We have a 4d SU(N) gauge theory,  on length scales larger than $R$,
 with an adjoint field $A_y$, and gauge coupling $g_4$ and very energetic domain walls.

What happens when $T$ goes up? First, at $T=T_d\sim\Lambda_{QCD}$ 4d SU(N)
deconfines. This means in the Euclidean version of the theory, that {\it thermal} Polyakov loop will acquire a VEV. A second, thermal, Z(N) symmetry is broken
too. In this deconfined phase there are still domain walls, because  $P(A_y)$
has still a VEV. The effective theory in this phase has a 3d SU(N) theory and two Higgs fields $A_0$ and $A_y$. 

Now the effective gauge coupling is dimensionful: $g_3^2=g_4^2T$. This is also the scale of the non-perturbative effects in this phase.
Increasing $T$ beyond the scale M will necessitate corrections to the effective 3d  theory, the lowest corrections being due to the mass scale M. So at some
temperature $T_c$ we have $g_3^2(T_c)=g_4^2T_c=M$. At this critical temperature:
$$T_c\sim M/g_4^2,$$
we arrive in the hot 5d SU(N) phase with dimensionless coupling $g_5^2T_c$.

The numerical value of this dimensionless coupling at $T_c$ is $g_5^2M/g_4^2=O(1)$, so also in
the hot 5d phase we have strong coupling. But above $T_c$ the VEV of the spatial Polyakov loop is zero, and the walls have disappeared.

An alternative way of arguing is to consider drops of Z(N) domains. The lower the surface tension, the easier one can form drops. The  energy of a drop
, compared to the energy without the drop, is proportional to its area $A$
.  On the other hand, the entropy of a drop is proportional to its area as well.  The probability for a drop to occur is then $\exp{(\epsilon(T)-\sigma(T))A}$. For $T=0$ the entropy is low and so is the probability.
But at high $T$  the entropy behaves like $\epsilon(T)\sim T/M$, whereas 
$\sigma(T)\sim {1\over{g_4}}(T/M)^{1/2}$~\cite{laine}. And again, parametrically at $T_c\sim M/g_4^2$ entropy takes over, rendering the probability of a drop large and hence restoring the symmetry.

 So our walls become relevant in cosmology: they will be created in a network, as the universe cools down through $T_c$.

\section{Lattice simulations and the cut-off in 5d gauge theory}

Clearly lattice simulations are needed to pin down these parametric estimates. 
We have done so for $N=2$, and the results are shown in fig.2.
\begin{figure}[tb]



\centerline{\begin{minipage}[c]{6.4cm}
    \psfig{file=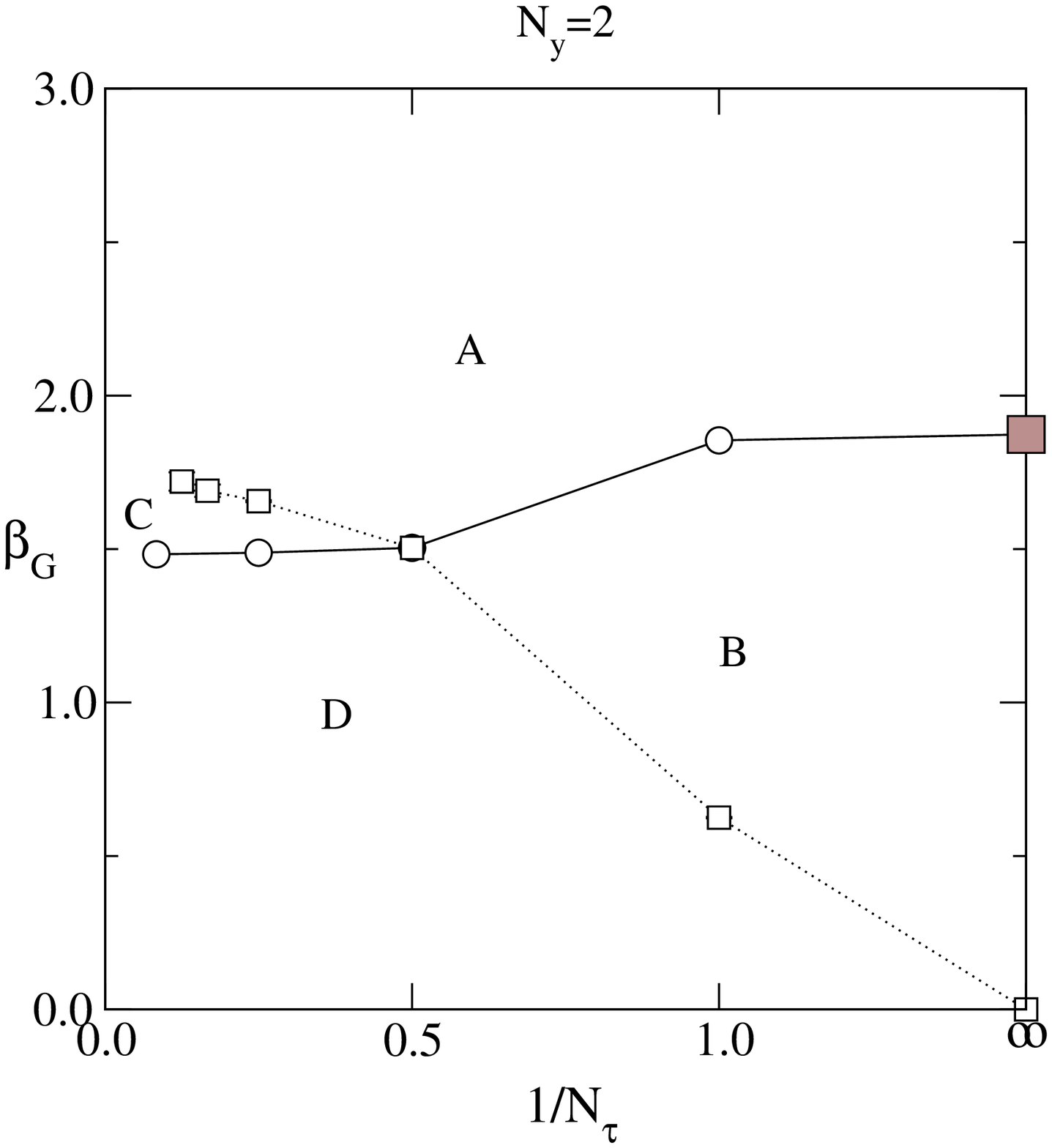,angle=0,width=6.4cm} \end{minipage}%
    ~~~\begin{minipage}[c]{6.4cm}
    \psfig{file=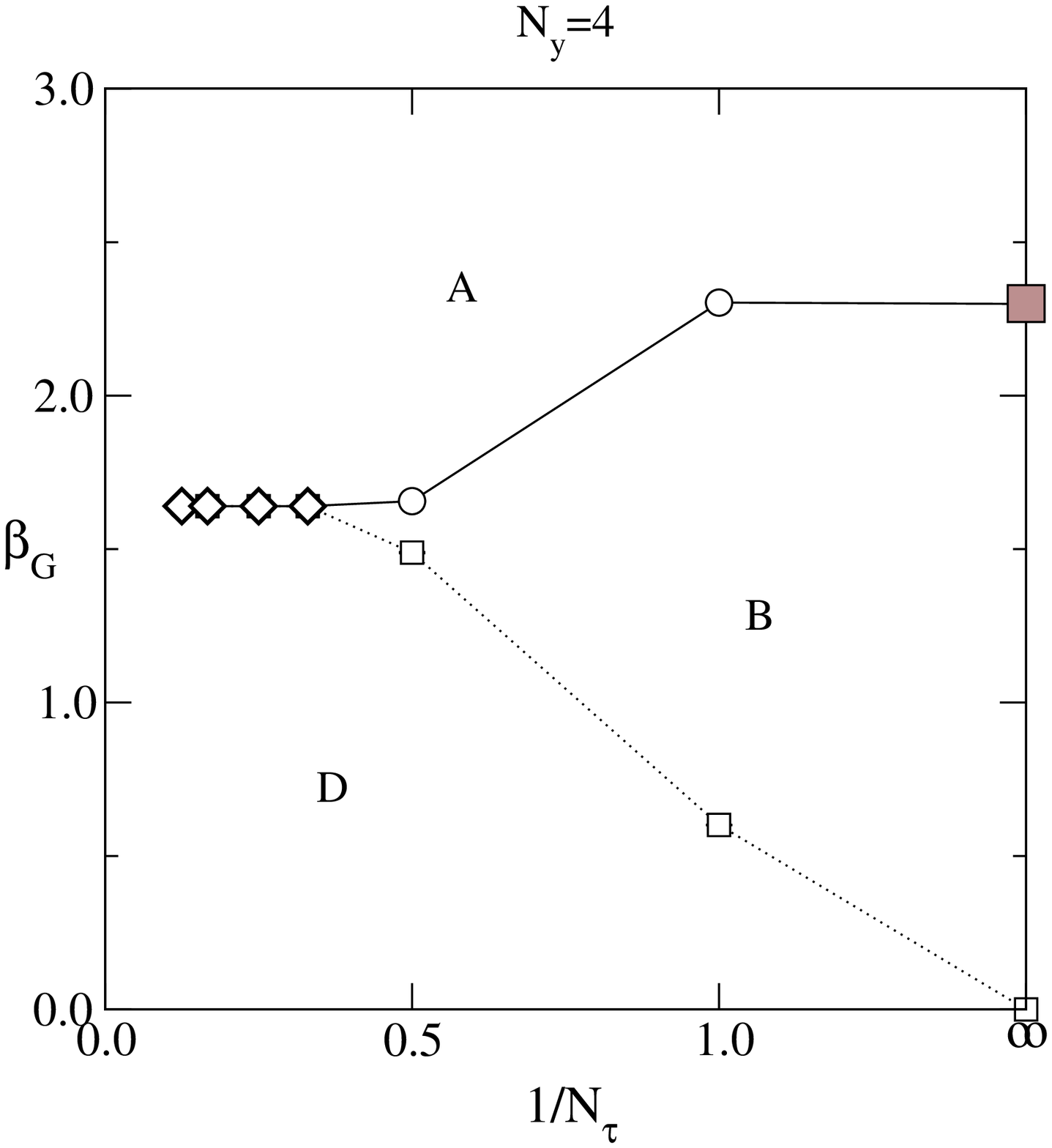,angle=0,width=6.4cm} \end{minipage}}

\vspace*{0.5cm}


\caption[a]{Phase diagrams for various $N_y$, in the 5d case. 
The lines connecting the simulation points are there to guide the eye
only.  Open circles and boxes denote second order
transitions, open diamonds first order ones.}

\end{figure}

Shown are simulations for a given value of the cut-off in every one of the four figures. In each we have gone to the 4d continuum limit as far as we could.
On the vertical axis the value of the coupling constant is plotted in units of the lattice cut-off $a$ : $\beta_G={4a\over{g_5^2}}$.

In the   figure on the left the four phases of fig.1 are clearly present.
 They are separated by lines where the thermal or spatial Polyakov loop does acquire a VEV. The transitions are everywhere smooth.

The right hand figure shows the cut-off in the fifth direction cannot become too small (lattice length $a\le R/4$). Then the phase with 4d physics at $T=0$ disappears altogether! The two smooth transition lines bordering this phase coalesce and the transition becomes first order. This is not unexpected~\cite{mc} from 5d bulk simulations.

As mentioned, the lattice parameter $\beta_G$ is related to the lattice cut-off $a$ and the 5d gauge coupling by  $\beta_G={4a\over{g_5^2}}$. So by multiplying the r.h.s. above and below by the scale $M$ one gets $\beta_G={4aM\over{g_4^2}}$.

The lattice parameter $N_y$ in the figure is related to the cut-off:
 $N_y={1\over{aM}}$. So ${1\over {g_4^2}}={N_y\over 4}\beta_G$ and has therefore an upper limit for $N_y=4$. T The point is that the presence of the cut-off does not allow $g_4^2$ to take arbitrary small values.

That means that $T_c=O(1){M\over {g_4^2}}$ has an upper limit $O(1)M$.

\section{Restrictions on cosmology}\label{sec:cosmo}

Clearly our wall network with energy density fluctuations on the order of
$(TeV)^4$ is not compatible with the CMB fluctuations.

Inflationary dynamics can dilute the network. Another possibility is to have 
the reheating temperature lower than our restoration transition temperature $T_c$, and we know $T_c$ is in the TeV region from the previous section. 
 
If one wants to identify the phase of the spatial Polyakov loop (a $J^{PC}=0^{+-}$ object in 4d) with the inflaton~\cite{inflaton}, the condition of slow roll gives a constraint:
\begin{equation}
g_4RM_{Pl}<<1,
\end{equation}
Our cut-off $a$ cannot be smaller than the Planck length, so, from our
lattice data on the cut-off, the size of our extra dimension should be  $R\sim M_{Pl}^{-1}\sim 1$, largely obeying the  constraint from CMB anisotropies~\cite{inflaton}.  Now there is no domain wall problem, since $T_c$ 
is above the Planck scale, and therefore no restored phase will occur.

Of course one can relax the constraints on the particle physics by admitting quarks in the bulk, and hence destabilize the walls. Or one can orbifold the extra dimension, implying that $A_y(y)=-A_y(R-y)$. Hence the constant mode is absent, and the Polyakov loop is constrained to be unity. By changing the topology of the  extra dimensions from $S^1\times S^1\times...$ one avoids Z(N) symmetry altogether.

\section{Acknowledgements}
\noindent I thank my collaborators for helpful comments.

\end{document}